\documentclass{emulateapj}
\usepackage[intlimits,fleqn]{amsmath}

\newcommand{\D}[1]{\, d #1 \,}
\newcommand{\vc}[1]{\vec{#1}}
\newcommand{\avg}[1]{\left< #1 \right>}
\newcommand{\ft}[1]{\widetilde{#1}}
\newcommand{\mcl}[1]{\mathcal{#1}}
\newcommand{\eps}{\varepsilon}
\newcommand{\vphi}{\varphi}

\shorttitle{Velocity Coordinate Spectrum}
\shortauthors{Chepurnov \& Lazarian}

\begin{document}

\title{Velocity Coordinate Spectrum: Geometrical Aspects of Observations}

\author{A. Chepurnov, A. Lazarian}
\affil{Department of Astronomy, University of Wisconsin, Madison, US}

\begin{abstract}
We analyze a technique of obtaining turbulence power spectrum using spectral line data along the velocity coordinate, which we refer to as Velocity Coordinate Spectrum (VCS). We formalize geometrical aspects of observation through a single factor, ``geometric term''. We find that all variety of particular observational configurations can be described using correspondent variants of this term, which we explicitly calculate. This allows us to obtain asymptotics for both parallel lines of sight and crossing lines of sight. The latter case is especially important for studies of turbulence within diffuse ISM in Milky Way. For verification of our results, we use direct calculation of VCS spectra, while the numerical simulations are presented in a companion paper. 
\end{abstract} 

\section{Introduction}

Turbulence is an essential property of interstellar medium, which controls a number of key astrophysical processes, including star formation, mass and heat transport and propagation of cosmic rays (see \citealt{Schlick99}, \citealt{Vaz00}, \citealt{NM01}, \citealt{Cho03}, see review by \citealt{CLV02}).  Observational studies of turbulence can put constraints on theoretical expectations for such processes and clarify their physical descriptions.

The advantage of obtaining the spectra of turbulence using the observations of Doppler broadened lines was obvious from the very beginning of the research in the field (see \cite{Laz06c} for a recent review). The problem happened to be extremely difficult, however.

Fortunately, we have seen some progress in this field recently. For instance, the VCA techniques for the analysis of turbulence from spectral data, developed by \cite{LP00} (henceforth LP00), is based on the analysis of fluctuations in channel maps. Those can be viewed as slices of a Position-Position-Velocity (PPV) cube. However, there exists a complimentary technique, employing statistics along velocity coordinate which has also been briefly discussed in that paper. This technique was introduced in the same (LP00) paper, but was not appreciated enough at the time of its introduction. In fact, it happened that studying the spectra of channel maps was an adopted procedure well before \cite{LP00} posed a question what these fluctuations may mean in terms of the underlying turbulence. On the contrary, has been studying the fluctuations of PPV emissivity along the velocity coordinate until very recently (see  \citealt{Che07}). As the result, the introduction of the technique was delayed. A detailed study of the technique,  that was termed Velocity Coordinate Spectrum (VCS) (Lazarian 2004), including the possible effect of absorption were presented in \cite{LP06} (henceforth LP06). However, there only a distant cloud is considered there as a possible turbulent volume for studies, which exclude the situation of not parallel lines of sight, which is typical, e.g. for studies of high latitude HI. 

In the present paper we present a formalism that in a concise form allows to reproduce the earlier results in the field, as well as to move the field forward. The gist of our approach is formalization of the role of observational geometry, which allows us to write out the basic equations of both the VCA and the VCS in a very similar way. Our approach also enables us to account for a limited resolution, various beam geometries and, more importantly, account for both the crossing or parallel lines of sight geometry of observations. In other words, unlike LP06  we can both consider turbulence studies within a distant cloud with size much larger than the distance to the cloud and the in volume that surrounds the observer. Moreover, our general formalism can easily be extended for the intermediate cases, when lines of sight are not crossing, but cannot be considered parallel either.

We note, that the VCS is special among the techniques of turbulence studies. Compared to the VCA or Velocity Centroids (see \cite{EL05} and references therein) it uses exclusively data along velocity coordinate. In this case the spatial information is used for improving statistics only. Our simulations in the companion paper confirm, that a very few independent measurements (e.g. $\sim 10$) is sufficient to get the underlying velocity turbulence spectrum with enough accuracy. This tool is irreplaceable when we do not resolve the object enough to analyze power spectrum in the pictorial plane. In addition, our present study makes the VCS the only tool accounting for convergence of lines of sight.

In what follows in Sect. \ref{sect:turb} we introduce the basic terms of statistics of turbulence, in Sect.  \ref{sect:SLine} we describe the signal at a spectrometer output in terms of underlying velocity and emissivity fields, in Sect. \ref{sect:CFS} we consider the basic statistics common for VCA and VCS, the power spectrum along the velocity coordinate is presented in Sect. \ref{sect:P1} and in Sect. \ref{sect:P1_asy} we consider its asymptotics. The variants of the geometric term are presented in Sect. \ref{sect:gc_all}. Some needed here facts regarding velocity and density statistics are presented in Appendices \ref{sect:vst} and \ref{sect:ecf}, the influence of the regular velocity shear is considered in Appendix \ref{sect:reg_vel}. We discuss our findings in Sect. \ref{sect:disc}.

\section{Underlying velocity and density turbulence} \label{sect:turb}

We assume here, that turbulent velocity and density are homogeneous and isotropic and admit a statistical description in terms of correlation (or structure) functions and corresponding power spectra. The latter pair of functions is related to each other through Fourier transform. 

For a vector field like velocity, correlation function and correspondent power spectrum take a tensor form. Power spectrum is different for solenoidal and potential field components. For each case spectrum's tensor properties can be represented by a factor, depending only on angles. Another, scalar, factor is responsible for energy distribution in the wavenumber space. 

Because of the self-similarity of turbulence, we usually expect a power law dependence for power spectrum. This power law can be either ``long wave dominated'' (or ``steep'') if the respective spectral index is greater than $3$, or ``short wave dominated'' (or ``shallow'') otherwise (see LP00). To keep total energy limited, both types of spectrum must have a cutoff: at low wavenumbers for the steep one, and at high wavenumbers for the shallow one. We assume the velocity spectrum to be steep, as the shallow velocity spectrum is not physically motivated. On the contrary, for density spectrum we consider both steep and shallow spectra. The shallow spectrum emerges for high Mach numbers (see \citealt{EL05}). 

The quantities we deal with in spectral line observations are velocity and emissivity. The latter can be proportional to density (like in emission lines of neutrals) or density squared (dielectron recombination lines in plasma). The regime of squared density modifies the underlying density spectrum as the emissivity spectrum is an auto-convolution of the density one in this case: the asymptotic slope of the emissivity spectrum is the same for steep density spectrum and different, $\alpha_\eps=2\alpha_\rho-3$, for the shallow one (see Appx. \ref{sect:ecf_st}).

We are interested in velocity statistics, and emissivity is considered only because it affects the  quantities, derived from observational data, that we use here. This influence is less for steep emissivity spectrum, and could be ignored (apart from the normalization), but for the shallow spectrum it should be accounted for. 

Because of non-linear character of the statistics we use Eq. [\ref{eq:P1_fin}], we can not directly recover the velocity spectrum. But if it is parametrized with the spectral index, spectrum amplitude, and cutoff wavenumber (related to the injection scale), we in principle can recover these parameters. 

\section{Spectral line signal at the spectrometer output} \label{sect:SLine}

The presentation of the techniques in both LP00 and LP06 starts from presenting the underlying turbulence statistics and proceeds with introducing the observables. In such presentation of the VCA and the VCS decouple rather early and the derivations are rather lengthy. Below we provide a derivation of the relevant formulae starting from the output of a spectrometer. This allows us to present both the VCA and the VCS within a unified approach. Moreover, it allows us to cover the observationally important case of crossing lines of sight. However, unlike LP06, we do not consider self-absorption, which means that our approach is applicable either to optically thin emission lines or to weak absorption lines.

\subsection{Emission line} \label{sect:SLine_em}

Let us write out space-velocity distribution of density:
\begin{equation} \label{eq:dens}
n_0(\vc{r},v)
  = n_0(\vc{r}) \varphi(v-v_{macro}(\vc{r}))
\end{equation}
\noindent where $\varphi(v)$ is Maxwellian distribution
\begin{equation} \label{eq:maxw}
\varphi(v)
  = \frac{1}{\sqrt{2\pi\beta}} \exp\left(-\frac{v^2}{2\beta}\right), \; \beta = \frac{k_B T}{m}
\end{equation}
\noindent and $n_0(\vc{r})$ is an integral density. 

If we write out the similar expression for emissivity 
\begin{equation} \label{eq:emiss}
\eps_0(\vc{r},v)
  = \eps_0(\vc{r}) \varphi_{\eps}(v-v_{macro}(\vc{r}))
\end{equation}
\noindent we will get
\noindent
$\eps_0(\vc{r}) = \gamma_1 n_0(\vc{r}), \; \vphi_\eps(v) = \varphi(v)$
-- for linear emissivity law,
\noindent
$\eps_0(\vc{r})=\gamma_2 n_0^2(\vc{r}), \; \vphi_\eps(v)=2\sqrt{\pi\beta} \varphi^2(v)$
-- for quadratic emissivity law,
\noindent where $\gamma_1$, $\gamma_2$ set up the ratio between the quantities integrated over the velocity.

Then we can write out the following expression for the signal in a spectrometer channel centered at the velocity $v_0$:
\begin{equation} \label{eq:sig}
S(v_0)
  = \frac{\lambda^3}{8\pi k_B} 
  \int \frac{ w_b(\phi,\theta) }{r^2} \D{\vc{r}} \eps_0(\vc{r}), 
  f(v_r(\vc{r}) + v_r^{reg}(\vc{r}) - v_0)
\end{equation}

\noindent where
\begin{equation} \label{eq:effsens}
f(v) =
  \int_{-\infty}^\infty \varphi_\eps(v'+v) f_s(v') \D{v'}
\end{equation}
In Eq. [\ref{eq:effsens}]
$f_s(v)$ is the channel sensitivity function, 
$w_b(\phi,\theta)$ is the instrument beam,
$\eps_0(\vc{r})$ is the integral emissivity,
$v_r(\vc{r})$ is the turbulent velocity radius-vector projection,
$v_r^{reg}(\vc{r})$ is the regular velocity radius-vector projection and
$\lambda$ is the wavelength of the spectral line.

In Eq. [\ref{eq:sig}] it is assumed that the following normalization is
 applicable: $\int w_b(\phi,\theta) \D{\Omega} = 1$ 
and $\int_{-\infty}^\infty f_s(v) \D{v} = 1$.

The principal quantity in our analysis, $S$ is measured in $K$, while $\eps_0$ is in $W/m^3$, $\lambda$ is in $m$ and $v$ is in $m/s$. We also assume that the instrument aperture efficiency is $1$. 

Random field $\eps_0(\vc{r})$ is not homogeneous, at least because the emitting structure is limited in space. To model this we introduce a homogeneous field $\eps(\vc{r})$ and a deterministic factor $w_\eps(\vc{r})$ setting up borders of an observed object. We will also ``pack'' into $\eps(\vc{r})$ all constant factors:
\begin{equation} \label{eq:eps0}
\eps_0(\vc{r})
  = \frac{8\pi k_B}{\lambda^3} w_\eps(\vc{r}) \eps(\vc{r}).
\end{equation}
It is also convenient to introduce a window function $w$ as follows:
\begin{equation} \label{eq:w}
w(\vc{r})
  \equiv \frac{1}{r^2} w_b(\phi,\theta) w_\eps(\vc{r}).
\end{equation}
With these notations we finally obtain:
\begin{equation} \label{eq:sig_rat}
S(v_0)
  = \int w(\vc{r}) \D{\vc{r}} \eps(\vc{r}) f(v_r(\vc{r}) + v_r^{reg}(\vc{r}) - v_0)
\end{equation}
In what follows we shall consider the Fourier transform of spectral line\footnote{Variable $k_v$ plays here the role of $k_z$ in LP00, being however different in dimension ($k_z = b k_v$). We use it here to avoid complications when $b=0$.}:
\begin{equation} \label{eq:sig_ft}
\begin{split}
\ft{S}(k_v) 
  & \equiv \frac{1}{2\pi} \int_{-\infty}^\infty S(v_0) e^{-i k_v v_0} \D{v_0} \\
  & = \ft{f}(k_v) \int w(\vc{r}) \D{\vc{r}} 
      \eps(\vc{r}) \exp(-i k_v (v_r(\vc{r}) + v_r^{reg}(\vc{r})))
\end{split}
\end{equation}
This function can be easily determined from observational data.

\subsection{Absorption line} \label{sect:SLine_abs}

\begin{figure}[p] 
\begin{center} 
\plotone{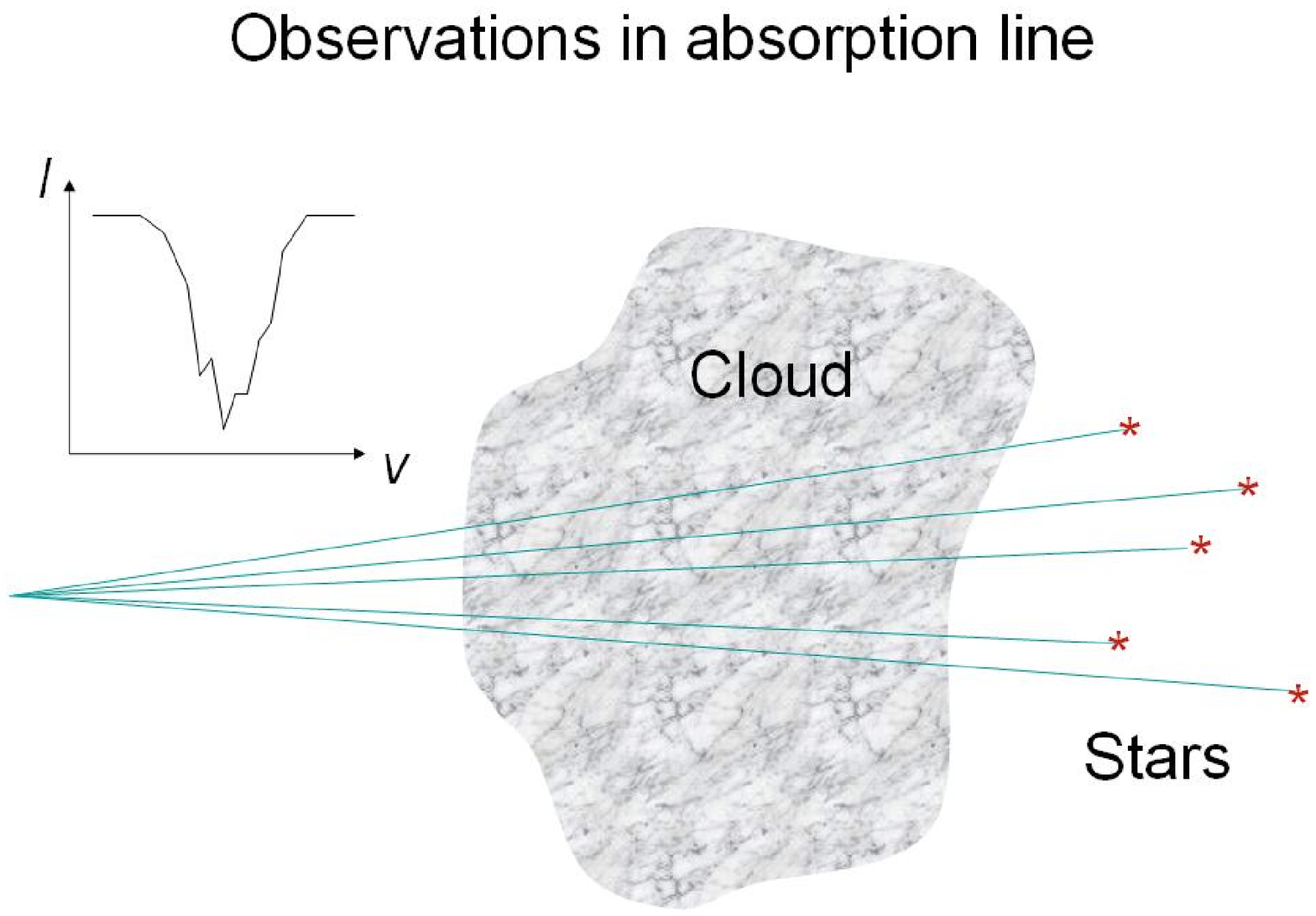}
\end{center}
\caption{VCS allows recovery of velocity statistics from absorption lines from stars. The whole $P_1$ is guaranteed to be in high-resolution mode, which provide better dynamical range over $k_v$. Numerical simulations show, that very few independent measurements are needed to gain enough statistics.\label{fig:abs}} 
\end{figure}

A signal in a channel $\D{v}$ received from a point object with the flux $F$ is as follows:
\begin{equation} \label{eq:dP_abs}
\begin{split}
\D{P}
  & = \frac{A_{eff} F}{2 k_B} \exp \left(-\alpha \int n_0(z,v)\D{z}\right) \D{v} \\
  & \approx \frac{A_{eff} F}{2 k_B} 
    \left(1 -\alpha \int n_0(\vc{r}) \varphi(v-v_{macro}(\vc{r})) \D{z}\right) \D{v}
\end{split}
\end{equation}
Then the signal can be written as an integral of $\D{P}$ weighted with the channel sensitivity function $f_s$:
\begin{equation} \label{eq:sig_abs}
\begin{split}
S_0(v_0)
  & = \frac{A_{eff} F}{2 k_B} \\
  & \cdot \left(1 -\alpha \int n_0(z) f(v(z) + v^{reg}(\vc{z}) - v_0) \D{z}\right)
\end{split}
\end{equation}
And, if we make the following substitution\footnote{as before, $n(z)$ is a homogeneous field and $w_n(z)$ is a correspondent window function}
\begin{equation} \label{eq:w_n}
  \frac{A_{eff} F \alpha}{2 k_B} n_0(z) \equiv w_n(z) n(z)
\end{equation}
and keep only the variable term of $S_0$ with inverted sign, we will have:
\begin{equation} \label{eq:sig_abs_rat}
S(v_0)
  \equiv \int w_n(z) \D{z} n(z) f(v_r(\vc{r}) + v_r^{reg}(\vc{r}) - v_0)
\end{equation}
The correspondent Fourier transform is:
\begin{equation} \label{eq:sig_abs_ft}
\ft{S}(k_v) 
  = \ft{f}(k_v) \int w_n(z) \D{z} 
    n(z) \exp(-i k_v (v(z) + v^{reg}(z)))
\end{equation}
The observational configuration for this regime is shown on Fig. \ref{fig:abs}.

\section{Common statistics for VCA and VCS} \label{sect:CFS}

Let us consider the following measure:
\begin{equation} \label{eq:K12_def} 
\begin{split}
K_{12}(k_v) 
  & \equiv \avg{\ft{S_1}(k_v) \ft{S_2}^*(k_v)} \\ 
  & =\ft{f}^2(k_v) \int w_1(\vc{r}) \D{\vc{r}} \int w_2(\vc{r'}) \D{\vc{r'}} \\
  & \cdot \avg{\eps(\vc{r})\eps(\vc{r'})} \avg{\exp(-i k_v (v_r(\vc{r})-v_{r'}(\vc{r'})))} \\
  & \cdot \exp(-i k_v (v_r^{reg}(\vc{r})-v_{r'}^{reg}(\vc{r'}))),
\end{split}
\end{equation} 
\noindent where indexes 1 and 2 designate different beam directions. We have assumed here that density and velocity are uncorrelated. The effects of such correlation have been studied analytically in LP00 and numerically in \cite{Laz01} and \cite{Esq03}. They were shown to be marginal.

The first averaging gives us an emissivity correlation function $C_\eps(\vc{r}-\vc{r'})$. Averaging of the exponent can be performed with assumption that the velocity field has Gaussian statistics:
\begin{equation*} 
\begin{split}
&\avg{\exp(-i k_v (v_r(\vc{r})-v_{r'}(\vc{r'})))} \\
&  = \exp\left(-\frac{k_v^2}{2} \avg{(v_r(\vc{r})-v_{r'}(\vc{r'}))^2}\right).
\end{split}
\end{equation*}

To proceed we assume that the beam separation as well as the beam width is small enough to neglect the difference between $v_r$ and $v_z$ (we consider $z$-axis to be a bisector of the angle between beams). We also assume that $v_z^{reg}(\vc{r})$ depends only on $z$ and admits a linear approximation:
\begin{equation} \label{eq:uz_lin} 
v_z^{reg}(z)
  = b(z-z_0)+v_{z,0}^{reg}.
\end{equation}

This leads us to
\begin{equation} \label{eq:K12_inter} 
\begin{split} 
K_{12}(k_v) 
  & =\ft{f}^2(k_v) \int w_1(\vc{r}) \D{\vc{r}} \int w_2(\vc{r'}) \D{\vc{r'}} C_\eps(\vc{r}-\vc{r'}) \\
  & \cdot \exp \left(-\frac{k_v^2}{2} D_z(\vc{r}-\vc{r'}) - ik_v b(z-z') \right), 
\end{split}
\end{equation}
\noindent where  $D_z$ is a velocity structure tensor projection:
\begin{equation} \label{eq:Dz_def} 
D_z(\vc{r}-\vc{r'}) 
  \equiv \avg{(v_z(\vc{r})-v_z(\vc{r'}))^2}, 
\end{equation}  
\noindent assuming that the velocity field is homogeneous.

Having substituted $\vc{r}-\vc{r'}$, we can write the following expression for $K_{12}$:
\begin{equation} \label{eq:K12_fin} 
\begin{split} 
K_{12}(k_v) 
  & =\ft{f}^2(k_v) \int w_{12}(\vc{r}) \D{\vc{r}} \\
  & \cdot C_\eps(\vc{r}) \exp \left(-\frac{k_v^2}{2} D_z(\vc{r}) - ik_v b z \right) 
\end{split}
\end{equation}
\noindent where
\begin{equation} \label{eq:w12_def} 
w_{12}(\vc{r}) 
  \equiv \int w_1(\vc{r'}) w_2(\vc{r'}+\vc{r}) \D{\vc{r'}} 
\end{equation} 
\noindent which will be further referred to as ``geometric term''. 

The case $w_{12}(\vc{r}) = \delta(\vc{R}-\vc{R_0}) w_{\eps,a}(z)$ leads us to the VCA\footnote{Parallel lines of sight geometry is assumed here. $\vc{R}$ is a 2-d vector in picture plane, $w_{\eps,a}$ is defined in Sect.\ref{sect:gc_irp}}, while making the two beams coincide, we turn to the VCS.

\section{Power spectrum along the velocity coordinate} \label{sect:P1}

Setting $w_1=w_2$ in Eq. [\ref{eq:K12_fin}] we can write:
\begin{equation} \label{eq:P1_fin} 
\begin{split} 
P_1(k_v)
  & \equiv K_{12}(k_v) |_{w_1=w_2} \\
  & =\ft{f}^2(k_v) \int w_a(\vc{r}) \D{\vc{r}} \\
  & \cdot  C_\eps(\vc{r}) \exp \left(-\frac{k_v^2}{2} D_z(\vc{r}) - ik_v b z \right) 
\end{split}
\end{equation}
\noindent where
\begin{equation} \label{eq:wa_def} 
w_a(\vc{r}) 
  \equiv w_{12}(\vc{r})|_{w_1=w_2} 
  = \int w(\vc{r'}) w(\vc{r'}+\vc{r}) \D{\vc{r'}} 
\end{equation} 

Some variants of $w_a$ are presented in Appendix \ref{sect:gc_all}. 

We get an important case of infinite resolution after applying of $\delta$-function in $w_a$ from Appendix \ref{sect:gc_irp}: 

\begin{equation} \label{eq:P1_ir_fin} 
\begin{split} 
P_1(k_v)
  & =\ft{f}^2(k_v) \int w_\eps(z) \D{z} \\
  & \cdot C_\eps(z) \exp \left(-\frac{k_v^2}{2} D_z(z) - ik_v b z \right) 
\end{split}
\end{equation}

In particular, it arises when using an absorption line [\ref{eq:sig_abs_rat}] from distant point sources behind the studied turbulent cloud (see Fig. \ref{fig:abs}).

Another important case arises when the emissivity spectrum is shallow and $C_\eps$ splits into constant and singular terms:
\begin{equation} \label{eq:Ce_split}
C_\eps(\vc{r}) 
  \approx \avg{\eps}^2 \left(1 + \left(\frac{r}{r_0}\right)^{3-\alpha_\eps}\right)
\end{equation}
\noindent and $P_1$ is splitted accordingly:
\begin{equation} \label{eq:P1_split} 
P_1(k_v) 
  = P_{1,v}(k_v) + P_{1,\eps}(k_v)
\end{equation}

In Eq. [\ref{eq:P1_split}] the first term with the exception of its amplitude is entirely determined by velocity field, while the other one depends on both emissivity and velocity.

\section {Geometric term} \label{sect:gc_all}

In this section we consider geometric terms needed for specific variants of $P_1$. The cases presented below roughly cover all possible geometrical configurations, which include on the instrument beam shape and relative location of the observed turbulent structure. Each of these variants of geometric term has different behavior for $r \to 0$ (see Tab. \ref{tab:gc}), which results in different slopes for $P_1$ asymptotics, even if velocity and emissivity statistics are the same (see Tab. \ref{tab:P1v_asy} and Tab. \ref{tab:P1eps_asy}).

\subsection{Ideal resolution, pencil beam} \label{sect:gc_irp}

Let $\vc{R}$ be a vector in a plane, perpendicular to $z$-axis. Assuming $\theta \approx \frac{R}{z}$ we have:
\begin{equation*}
\int w_b\left(\phi,\frac{R}{z}\right) \D{\vc{R}} = z^2
\end{equation*}
So we can model ``infinitely narrow'' beam as
\begin{equation} \label{eq:wb_irp}
w_b(\vc{r}) 
  = z^2 \delta(\vc{R})
\end{equation}

Then the window function is as follows:
\begin{equation} \label{eq:w_irp}
w(\vc{r})
  \approx \frac{1}{z^2} w_b(\vc{r}) w_\eps(z)
  = \delta(\vc{R}) w_\eps(z)
\end{equation}
\noindent and we can write the expression for $w_a$:
\begin{equation} \label{eq:wa_irp}
w_a(\vc{r})
  = \delta(\vc{R}) w_{\eps,a}(z)
\end{equation}
\noindent where
\begin{equation} \label{eq:w_eps_a_z}
w_{\eps,a}(z)
  = \int_{-\infty}^\infty w_\eps(z') w_\eps(z'+z) \D{z'}
\end{equation}

Besides ``very thin'' pencil beam, this case is relevant for observation of turbulence in an absorption line of a remote point source (see Fig. \ref{fig:abs}).

\subsection{Limited resolution, parallel lines of sight} \label{sect:gc_plos}

In this case emitting structure is far enough, so that the lines of sight do not converge within beam size. Therefore the corresponding window function $w_\eps$ is non-zero only in some vicinity of a distant point $z_0$. We also assume, that picture-plane extent of the object is bigger enough than the beam size, so the dependence on $\vc{R}$ of $w_\eps$ can be ignored.

Then we have:
\begin{equation*}
w(\vc{r})
  \approx \frac{1}{z_0^2} w_b(\theta) w_\eps(z)
\end{equation*}

If we introduce
\begin{equation} \label{eq:Wb_plos_def}
W_b(R)
  \equiv \frac{1}{z_0^2} w_b\left(\frac{R}{z_0}\right),
  \qquad \smallint W_b(R) \D{\vc{R}} = 1
\end{equation}
\noindent window function becomes
\begin{equation*}
w(\vc{r})
  = W_b(R) w_\eps(z)
\end{equation*}
\noindent and for $w_a$ we can write:
\begin{equation} \label{eq:wa_plos}
w_a(\vc{r})
  = W_{b,a}(R) w_{\eps,a}(z)
\end{equation}
\noindent where
\begin{equation} \label{eq:Wba_plos}
W_{b,a}(R)
  = \int W_b(\vc{R'}) W_b(\vc{R'}+\vc{R}) \D{\vc{R'}}
\end{equation}

\subsection{Limited resolution, crossing lines of sight} \label{sect:gc_clos}

Here we consider the observation point to be inside or near the emitting structure. We also assume, that $w_\eps$ depends only on $z$. 

For our estimations we have to choose some particular form of $w_b$. For calculation convenience we may take the Gaussian: 
\begin{equation} \label{eq:wb_gau}
w_b(\theta)
  = \frac{1}{\pi\theta_0^2} e^{-\frac{\theta^2}{\theta_0^2}}
\end{equation}

If we introduce
\begin{equation} \label{eq:Wb_clos_def}
W_b(\vc{r})
  \equiv \left\{ 
    \begin{array}{ll} 
      \frac{1}{z^2} w_b\left(\frac{R}{z}\right)
      = \frac{1}{\pi\theta_0^2 z^2} e^{-\frac{R^2}{\theta_0^2 z^2}}, & z>0 \\ 
      0, & z<0
    \end{array} 
  \right. 
\end{equation}
\noindent the window function takes the following form:
\begin{equation*}
w(\vc{r})
  = W_b(\vc{r}) w_\eps(z)
\end{equation*}
\noindent and after some algebra we have:
\begin{equation} \label{eq:wa_clos}
\begin{split}
w_a(\vc{r}) 
  & = \frac{1}{\pi \theta_0^2} \int_0^\infty \frac{\D{z'}}{z'^2+(z'+|z|)^2} \\
  & \cdot \exp \left( -\frac{R^2}{\theta_0^2 (z'^2+(z'+|z|)^2)}\right) \\
  & \cdot w_\eps(z') w_\eps(z'+|z|) 
\end{split}
\end{equation}

\subsubsection{Asymptotics}

As this function is singular at $r=0$, we also need its asymptotic behavior. If $z_{edge}$ is a characteristic scale of $w_\eps$, we can assume
\begin{equation} \label{eq:R_z_assump}
z \ll z_{edge}, \quad R \ll \frac{\theta_0 z_{edge}}{2}
\end{equation}
\noindent and set $w_\eps=1$. Then, if we set $R=0$, we have:
\begin{equation} \label{eq:wa_clos_z}
w_a(0,z)
  = \frac{1}{4 \theta_0^2 z}
\end{equation}

Alternatively, setting $z=0$ we have:
\begin{equation} \label{eq:wa_clos_R}
w_a(R,0) 
  = \frac{1}{2\sqrt{2\pi}\theta_0 R}
\end{equation}

Approximating $\theta$-dependence with an ellipse, we finally obtain:
\begin{equation} \label{eq:wa_clos_asy}
\begin{split}
w_a(\vc{r})
  & \approx \frac{1}{\sqrt{8 \pi \theta_0^2 R^2 + 16 \theta_0^4 z^2}} \\
  & = \frac{1}{2 \theta_0 r \sqrt{2 \pi \sin^2 \theta + 4 \theta_0^2 \cos^2 \theta}}
\end{split}
\end{equation}

\subsubsection{Approximation corresponding to tophat window function}

Equation [\ref{eq:wa_clos}] can be analytically estimated, if we take a tophat approximation for $w_\eps$:
\begin{equation} \label{eq:weps_tophat}
w_\eps(z)
  = \left\{ 
    \begin{array}{ll} 
      1, & z \in [z_0,z_1]\\ 
      0, & z \notin [z_0,z_1]
    \end{array} 
  \right. 
\end{equation}
\noindent where $z_0$ and $z_1$ define the borders of the observed structure. It gives us the following expression for the geometric term:
\begin{equation} \label{eq:wa_clos_appr}
\begin{split}
w_a(\vc{r}) 
  & \approx \dfrac{-1}{2\sqrt{\pi} \theta_0 R z} \cdot
    \dfrac{
      \arctan\left(1+\frac{2 z_0}{z}\right)+\arctan\left(1-\frac{2 z_1}{z}\right)
    }{
      \left(2 z_0^2 + p z^2\right)^{-\frac{1}{2}} - 
      \left(2 z_1^2 - p z^2\right)^{-\frac{1}{2}} 
    } \\
  & \cdot \left( 
      \mathrm{erf}\left(\dfrac{R}{\theta_0 \sqrt{2 z_0^2 + p z^2}} \right) - 
      \mathrm{erf}\left(\dfrac{R}{\theta_0 \sqrt{2 z_1^2 - p z^2}} \right)
    \right) 
\end{split}
\end{equation}
\noindent where
\begin{equation*} 
  p
    = \dfrac{z_1+z_0}{z_1-z_0}
\end{equation*}

This case is applicable if the observation point is outside, but still near to the emitting structure. This way we can account for the Local Bubble when observing the nearby turbulence in the Milky Way. 

\subsection{Flat beam, parallel lines of sight} \label{sect:gc_irfp}

Flat beam can be expressed as follows:
\begin{equation} \label{eq:wb_irfb}
w_b(\theta,x) = z w_{b,\theta}(\theta) \delta(x)
\end{equation}

In the case of remote emitting structure the corresponding window function $w_\eps$ is non-zero only in some vicinity of a distant enough point $z_0$. Then the window function $w$ is as follows:
\begin{equation} \label{eq:w_irfb}
w(\vc{r})
  \approx \frac{1}{z_0} w_{b,\theta}(\frac{y}{z_0}) \delta(x) w_\eps(z)
  \equiv W_b(\vc{R}) w_\eps(z)
\end{equation}

\noindent Then
\begin{equation} \label{eq:Wba_irfb}
W_{b,a}(\vc{R})
  = \frac{\delta(x)}{z_0} 
    \int_{-\pi}^\pi w_{b,\theta}(\theta) w_{b,\theta}(\theta+\frac{y}{z_0}) \D{\theta}
  \equiv \delta(x) W_{b,\theta,a}(y)
\end{equation}
\noindent and
\begin{equation} \label{eq:wa_irfb}
w_a(\vc{r})
  = \delta(x) W_{b,\theta,a}(y) w_{\eps,a}(z)
\end{equation}

\subsection{Flat beam, crossing lines of sight} \label{sect:gc_irfc}

In this case the window function has the following form: 
\begin{equation} \label{eq:w_irfc}
w(\vc{r})
  \approx \frac{1}{z} w_{b,\theta}(\frac{y}{z}) \delta(x) w_\eps(z)
\end{equation}

\noindent If we set
\begin{equation} \label{eq:wbth_irfc}
w_{b,\theta}(\theta) = \frac{1}{\sqrt{\pi} \theta_0} e^{-\frac{\theta^2}{\theta_0^2}}
\end{equation}
\noindent and
\begin{equation} \label{eq:weps_gauss}
w_\eps(z)
  \equiv \left\{ 
    \begin{array}{ll} 
      e^{-\frac{z^2}{z_{edge}^2}}, & z>0 \\ 
      0, & z<0
    \end{array} 
  \right. 
\end{equation}
\noindent we will have
\begin{equation} \label{eq:wa_irfc}
\begin{split}
w_a(\vc{r}) 
  & = \frac{\delta(x)}{\sqrt{\pi} \theta_0} \int_0^\infty \frac{\D{z'}}{\sqrt{z'^2+(z'+|z|)^2}} \\
  & \cdot \exp \left( -\frac{y^2}{\theta_0^2 (z'^2+(z'+|z|)^2)}-\frac{z'^2+(z'+|z|)^2}{z_{edge}^2}\right)
\end{split}
\end{equation}

If we set $y=0$, for $z \approx 0$ we have:
\begin{equation*}
w_a(x,0,z) 
  \approx - \frac{\delta(x)}{\sqrt{2\pi} \theta_0} \log \left( \frac{z}{z_{edge}} \right)
\end{equation*}

Otherwise, if $z=0$, for $y \approx 0$ we get:
\begin{equation*}
w_a(x,0,z) 
  \approx - \frac{\delta(x)}{\sqrt{2\pi} \theta_0} \log \left( \frac{y}{\theta_0 z_{edge}} \right)
\end{equation*}

Approximating dependence on $\theta$ with an ellipse, we get:
\begin{equation} \label{eq:wa_appr_irfc}
w_a(\vc{r}) 
  \approx - \frac{\delta(x)}{\sqrt{2\pi} \theta_0} 
  \log \left( \frac{r}{z_{edge}} \sqrt{\frac{1}{\theta_0} \sin^2 \theta + \cos^2 \theta }\right)
\end{equation}

\section{Asymptotics of \protect{$P_1$}} \label{sect:P1_asy}

The parameters of high-$k_v$ asymptotics of $P_1$ for different emissivity spectrum types and geometries are presented in Tab. \ref{tab:P1v_asy}, Tab. \ref{tab:P1eps_asy}, Tab.\ref{tab:Au} and Tab.\ref{tab:Aeps}. Here ``high'' actually means ``big enough for taking asymptotics for all remaining terms in Eq. [\ref{eq:P1_fin}]''. Velocity structure tensor projection $D_z$ in the exponent term is replaced by its asymptotics Eq. [\ref{eq:Dz_r_asy}] as well. Because of suppression by this term, other terms are either factored out, if they are continuous at zero, or remain in the integrand, if singular. 

In particular, emissivity correlation function $C_\eps$ is continuous for the steep spectrum and singular for the shallow one (see Appendix \ref{sect:ecf_st}). Geometric term $w_a$ is singular in the regimes of infinite resolution (2-d $\delta$-function), crossing lines of sight ($r^{-1}$), flat beam with parallel lines of sight (1-d $\delta$-function), flat beam with crossing lines of sight (1-d $\delta$-function and logarithmic singularity), and continuous for parallel lines of sight in low-resolution mode (see Sect. \ref{sect:gc_all}). In general we can say, that for more prominent singularity in the integrand we have shallower $P_1$.


The asymptotic regime of $P_1$ depends on $k_v$: with the real finite-size beam after some transition point we switch from the high-resolution to the low-resolution mode with steeper slope. 

This point corresponds to the situation when the exponent argument in Eq. [\ref{eq:P1_fin}] in transverse direction is about unity at beam radius $z_0 \theta_0$. This gives us (for the parallel l.o.s.):
 \begin{equation} \label{eq:kv_t_plos} 
k_{v,t} 
  = \frac{1}{\sqrt{2\pi I_s V_0^2} (z_0 \theta_0)^{\frac{\alpha_v-3}{2}}} 
\end{equation} 

For crossing l.o.s. with account for Eq. [\ref{eq:wa_clos_R}] we have:
\begin{equation} \label{eq:kv_t_clos} 
k_{v,t} 
  = \frac{1}{
    \sqrt{2\pi I_s V_0^2} (\frac{z_{edge}}{2} \theta_0)^{\frac{\alpha_v-3}{2}}
  } 
\end{equation}

It can be shown, that this is equivalent to
\begin{equation} \label{eq:kv_gen} 
k_{v,t} 
  = \left(
    \int \D{\Omega} \int_{\frac{1}{r_0}}^\infty k^2 \D{k} U_{zz}^2(\vc{k})
  \right)^{-1}
\end{equation}
\noindent where $r_0 = z_0 \theta_0$ for parallel and $r_0 = z_{edge} \theta_0 /2$ for crossing lines of sight.
The latter expression is simply $k_v$, which corresponds to the velocity variance at the beam scale. 

\begin{figure}[p] 
\begin{center} 
\plotone{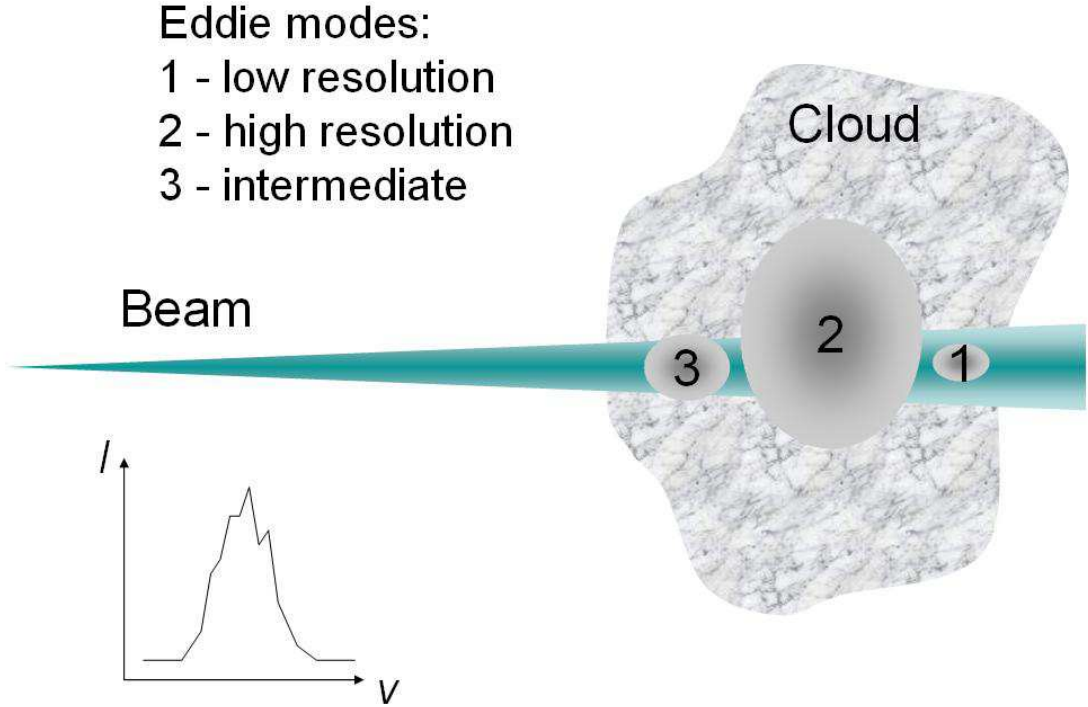}
\end{center}
\caption{Illustration to VCS resolution regimes. Eddies within beam size are in low resolution mode and cause steepening of $P_1$. The ones exceeding the beam size are in high-resolution mode and correspondent $P_1$ is shallower. Therefore high-resolution mode is more preferable regarding to the signal-to-noise ratio.\label{fig:res}} 
\end{figure}

In Fig. \ref{fig:res} we illustrate the resolution regimes of the VCS technique. Depending on the beam size one may or not resolve the spatial extent of the eddies under study.

\begin{figure}[p] 
\begin{center} 
\plottwo{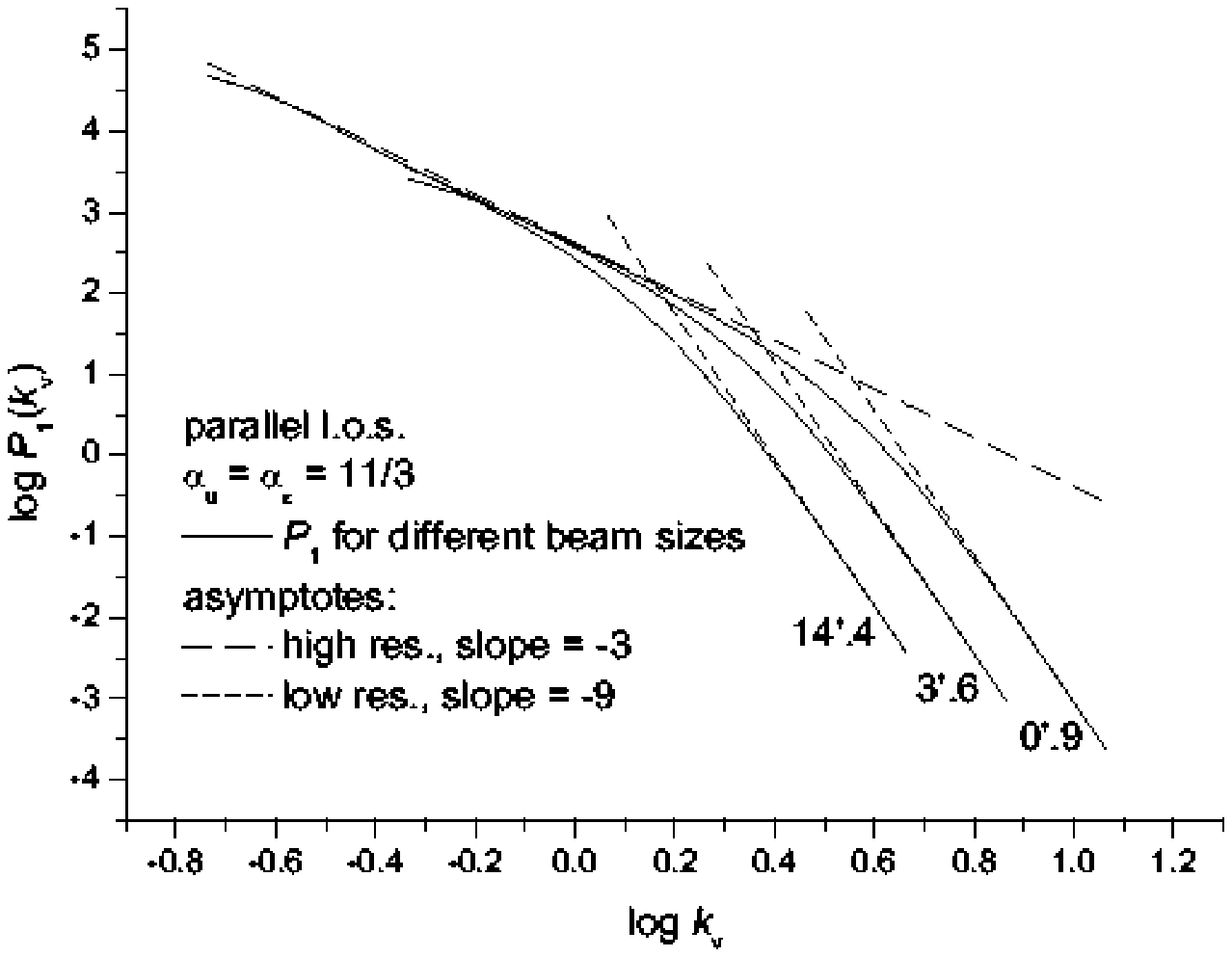}{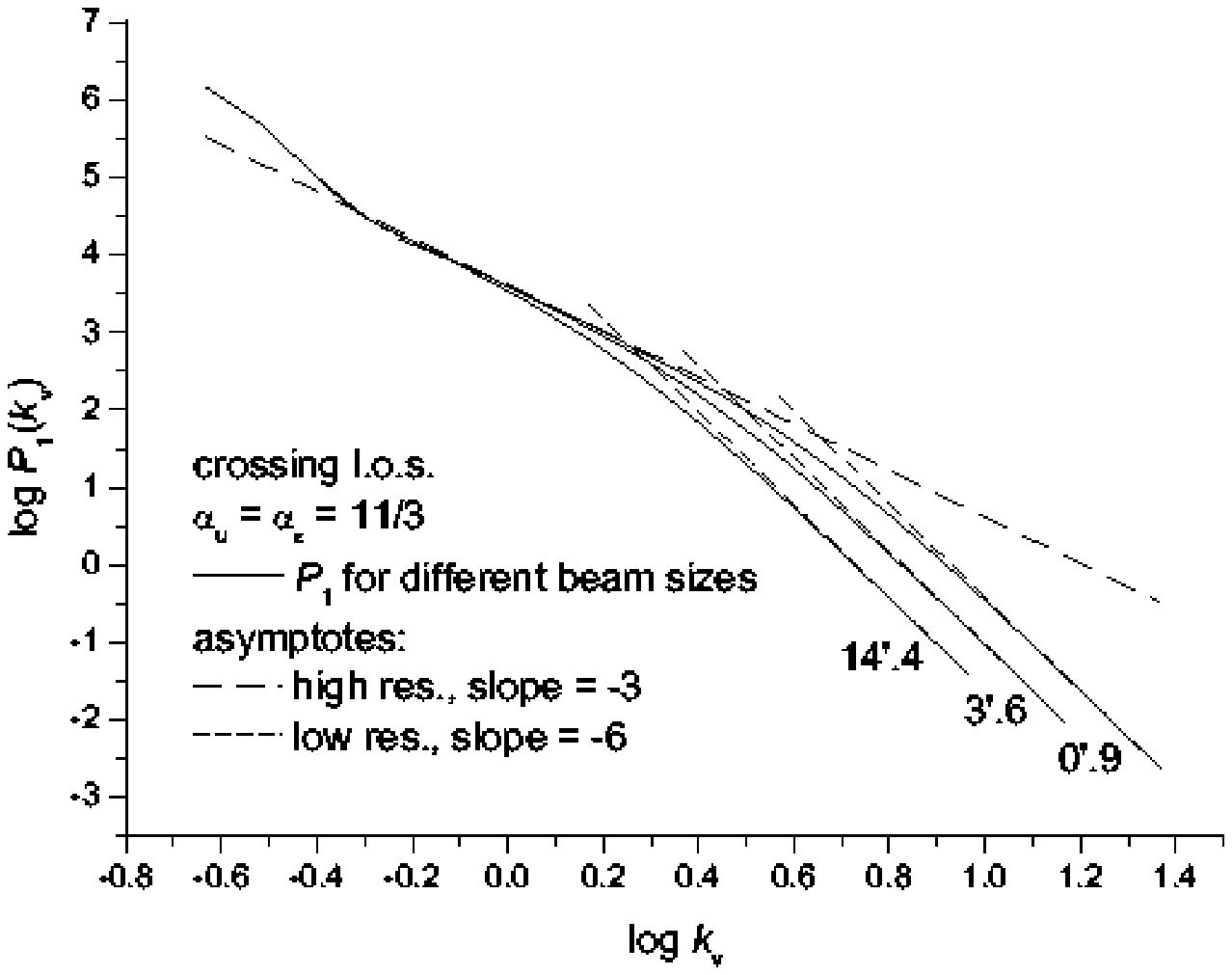}
\end{center}
\caption{One-dimensional power spectrum along velocity coordinate (direct calculation). Left: parallel lines of sight, distance to the object is 1000, object thickness is 100, injection scale is 30. Right: observation point is inside the emitting structure (converging lines of sight), distance to emitting structure's edge is 1000, injection scale is 30.\label{fig:P1_calc}} 
\end{figure}

To check the obtained results for asymptotical regimes we have performed direct calculation of the primary expression of $P_1$ [\ref{eq:P1_fin}]. The results presented on Fig. \ref{fig:P1_calc} confirm our estimations of asymptotic amplitudes and slopes. Numerical verifying with synthetic fields is discussed in detail in \cite{LC07}.

\section{Discussion} \label{sect:disc}

In the paper above we have presented the VCS technique in a form that, first of all, reveals the close connection between the VCS and the VCA. More importantly, we have described a new case very important for studies of galactic HI, namely, the case of collecting signal along converging lines of sight. 

We can change the effective instrument resolution to switch between regimes of $P_1$. This could be used as an additional test, if the chosen model is adequate. In principle, the relation between $P_1$ amplitudes in high and low resolution regimes can give us additional information about ISM, such as compressibility. Further research should show how good must be the data to enable us to do such a study.

It can happen that the asymptotic regime is not achieved because of thermal smoothing of the spectral line, or low spectrometer resolution. In this case we can calculate velocity field parameters by fitting of the model $P_1$ to the data with different spacial smoothing. This approach is adopted, for instance, in \cite{Che07}.

\begin{figure}[p] 
\begin{center}
\plotone{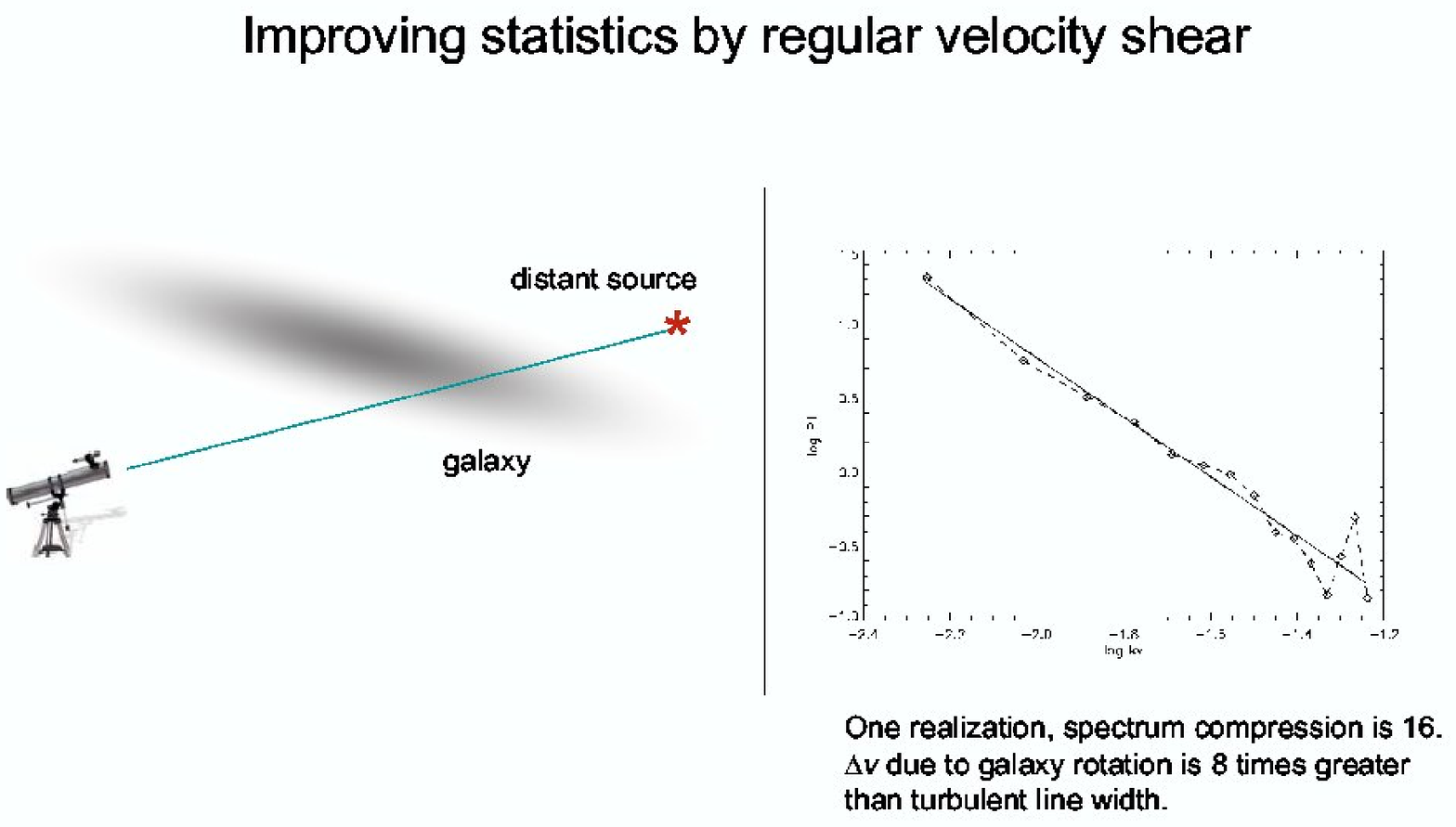}
\end{center}
\caption{VCS can be applied even to a single spectral line. Configuration can be as shown on this picture. Regular velocity shear from the galactic rotation results in statistics being sufficient to recover $P_1$\label{fig:shear}} 
\end{figure}

As mentioned above, VCS is a natural choice if we do not have an adequate spatial information. It can be applied to observations of absorption lines in spectra of a number of distant sources behind the cloud under study (Fig. \ref{fig:abs}). Our simulations in a companion paper show, that the required number of independent measurements could be only $\sim 10$. The spectrum $P_1$ is in high-resolution mode in this case, see Eq. [\ref{eq:P1_ir_fin}]. In some cases we may be using the  spectrum measured along only one line of sight, but use regular velocity shear to increase the effective statistical sample. In this case we ``unfold'' the velocity fluctuations and this leads to reacher statistics, that we can utilize by averaging of adjacent $k_v$-harmonics, see Fig. \ref{fig:shear}.

Another important case for using the VCS can be the studies of turbulence in galaxy clusters \cite{Vikh98}. Such studies can utilize X-rays, but in most cases the spatial resolution of telescopes may not be adequate to get channel maps with a sufficient spatial resolution.

As discussed earlier in LP06, the inertial range observed with the VCS may be somewhat limited. Some extension is possible by using heavier species, for which the thermal broadening is reduced. Another way that we explore in practical terms in \cite{Che07} is fitting the data with a model and using the VCS formulae directly, rather than their power-law asymptotics. Naturally, the VCS is a complementary technique to the VCA and their use can improve the reliability of interpreting fluctuations in the shapes of Doppler broadened lines in terms of underlying spectra of turbulence.

\appendix

\section{Velocity structure tensor} \label{sect:vst}

In order to get the result normalized with respect to spectrum parameters we express the structure tensor projection $D_z$ in terms of spectral tensor. 
 
If we consider $\vc{v}(\vc{r})$ solenoidal with power-law power spectrum having cutoff at large scales, spectral tensor will be as follows (\citealt{Les91}):
\begin{equation} \label{eq:Fu}
F_{ij}(\vc{k}) 
  = \frac{V_0^2}{k^{\alpha_v}} e^{-\frac{k_0^2}{k^2}}
  \left(\delta_{ij} - \frac{k_i k_j}{k^2}\right)
\end{equation}
\noindent where
$V_0^2$ is a velocity power spectrum amplitude,
$\alpha_v$ is a velocity spectral index and
$k_0$ is a cutoff wavevector.

Then $D_z$ is as follows:
\begin{equation} \label{eq:Dz_Fu}
D_z(\vc{r}) 
  = 2 \int \D{\vc{k}} (1 - e^{i\vc{k}\vc{r}}) \hat{z}_i\hat{z}_j F_{ij}(\vc{k})
\end{equation}
After performing the integration over wavevector directions we will get: 
 \begin{equation} \label{eq:Dz_r} 
D_z(r,\theta) 
  = 4\pi V_0^2 r^{\alpha_v-3} (I_c(r) \cos^2 \theta + I_s(r) \sin^2 \theta) 
\end{equation} 
\noindent where $I_c(r)$ and $I_s(r)$ are responsible for saturation of $D_z$ at large $r$'s: 
\begin{equation} \label{eq:Ic} 
I_c(r) 
  = \frac{4}{3} 
    \int_0^\infty \frac{\D{q}}{q^{\alpha_v-2}}\exp \left( -(k_0 r)^2/q^2 \right) \cdot 
    \left(1-\frac{3}{q^2}\left(\frac{\sin q}{q} - \cos q \right) \right) 
\end{equation} 
\begin{equation} \label{eq:Is}
\begin{split} 
I_s(r)
  & =  
    2 \int_0^\infty \frac{\D{q}}{q^{\alpha_v-2}}\exp \left( -(k_0 r)^2/q^2 \right) \cdot 
    \left( 1-\frac{\sin q}{q} \right) \\ 
  & - \frac{2}{3} 
    \int_0^\infty \frac{\D{q}}{q^{\alpha_v-2}}\exp \left( -(k_0 r)^2/q^2 \right) \cdot 
    \left(1-\frac{3}{q^2}\left(\frac{\sin q}{q} - \cos q \right) \right) 
\end{split}
\end{equation} 

If $r \sim 0$, we can use 
\begin{equation} \label{eq:Dz_r_asy} 
D_z(r,\theta)
  \approx 4\pi V_0^2 r^{\alpha_v-3} (I_c \cos^2 \theta + I_s \sin^2 \theta) 
\end{equation} 
\noindent (in this text $I_*$ without argument means $I_*(0)$)

For potential field we have:
\begin{equation} \label{eq:Fv_p}
F_{ij}(\vc{k}) 
  = \frac{V_0^2}{k^{\alpha_v}} e^{-\frac{k_0^2}{k^2}}\frac{k_i k_j}{k^2}
\end{equation}

In this case
\begin{equation} \label{eq:Ic_p} 
\begin{split} 
I_c(r) 
  & =  
    2 \int_0^\infty \frac{\D{q}}{q^{\alpha_v-2}}\exp \left( -(k_0 r)^2/q^2 \right) \cdot 
    \left( 1-\frac{\sin q}{q} \right) \\ 
  & - \frac{4}{3} 
    \int_0^\infty \frac{\D{q}}{q^{\alpha_v-2}}\exp \left( -(k_0 r)^2/q^2 \right) \cdot 
    \left(1-\frac{3}{q^2}\left(\frac{\sin q}{q} - \cos q \right) \right) 
\end{split}
\end{equation} 
\begin{equation} \label{eq:Is_p}
I_s(r)
  =  
  \frac{2}{3} 
    \int_0^\infty \frac{\D{q}}{q^{\alpha_v-2}}\exp \left( -(k_0 r)^2/q^2 \right) \cdot 
    \left(1-\frac{3}{q^2}\left(\frac{\sin q}{q} - \cos q \right) \right) 
\end{equation} 

\section{Emissivity correlation function} \label{sect:ecf}

\subsection{Steep spectrum} \label{sect:ecf_st}

If spectral index $\alpha_\eps > 3$, the spectrum is long wave dominated and should have cutoff at low $k$'s to avoid divergence. If we assume an isotropic field, it may be as follows:
\begin{equation} \label{eq:Fe_st}
F(\vc{k}) 
  = \frac{\mcl{E}_0^2}{k^{\alpha_\eps}} \exp \left(-k_0^2/k^2\right)
\end{equation}

In this case the correlation function of a variable component of emissivity is
\begin{equation} \label{eq:Ce_st}
C_{\eps,0}(\vc{r})
  \equiv C_\eps(\vc{r}) - \avg{\eps}^2
  = 4\pi \int_0^\infty k^2 \D{k} \frac{\mcl{E}_0^2}{k^{\alpha_\eps}} 
    \exp \left(-k_0^2/k^2\right) \frac{\sin k r}{k r}
\end{equation}

The asymptotic behavior at $r \sim 0$ is clear if we write $C_\eps$ as follows:
\begin{equation} \label{eq:Ce_st_asy}
C_{\eps,0}(r)
  = C_\eps(0) - 4\pi \mcl{E}_0^2 r^{\alpha_\eps - 3} I_{st}(r) 
\end{equation}

\noindent where
\begin{equation} \label{eq:I_st}
I_{st}(r)
  = \int_0^\infty \frac{\D{q}}{q^{\alpha_\eps-2}} \exp \left(-k_0^2 r^2/q^2\right)
    \left( 1 - \frac{\sin q}{q} \right)
\end{equation}

For small $r$'s we can replace $I_{st}(r)$ with $I_{st}(0)$. 

Here we omit the second term in Eq. [\ref{eq:Ce_st_asy}] for the final asymptotics of $P_1(k_v)$.

\subsection{Shallow spectrum} \label{sect:ecf_sh}

Short-wave dominated spectrum ($\alpha_\eps < 3$) must have cutoff at large $k$'s:
\begin{equation} \label{eq:Fe_sh}
F(\vc{k}) 
  = \frac{\mcl{E}_0^2}{k^{\alpha_\eps}} \exp \left(-k^2/k_1^2\right)
\end{equation}

The correspondent correlation function is
\begin{equation} \label{eq:Ce_sh}
C_{\eps,0}(\vc{r})
  = 4\pi \int_0^\infty k^2 \D{k} \frac{\mcl{E}_0^2}{k^{\alpha_\eps}} 
    \exp \left(-k^2/k_1^2\right) \frac{\sin k r}{k r}
\end{equation}
It can be rewritten as follows:

\begin{equation*}
C_{\eps,0}(r)
  = \frac{4\pi \mcl{E}_0^2}{r^{3-\alpha_\eps}} I_{sh}(r) 
\end{equation*}

\noindent where
\begin{equation} \label{eq:I_sh}
I_{sh}(r)
  = \int_0^\infty \frac{\D{q}}{q^{\alpha_\eps-2}} \exp \left(-q^2/k_1^2 r^2\right)
    \frac{\sin q}{q}
\end{equation}

Considering $k_1$ being high enough, and keeping in mind, that relevant $r$'s should significantly exceed\footnote{These values of $r$ can however be enough small to take $r \to 0$ asymptotics for other terms in Eq. [\ref{eq:P1_fin}]} $1/k_1$ , we can use the following approximation:
\begin{equation} \label{eq:Ce_sh_asy}
C_{\eps,0}(r)
  \approx \frac{4\pi \mcl{E}_0^2}{r^{3-\alpha_\eps}} I_{sh}^\infty 
\end{equation}
\noindent where
\begin{equation} \label{eq:I_sh_infty}
I_{sh}^\infty
  = \int_0^\infty \frac{\sin q}{q^{\alpha_\eps-1}} \D{q}
\end{equation}

\subsection{Quadratic emissivity} \label{sect:ecf_sq}

The expressions [\ref{eq:Fe_st}] and  [\ref{eq:Fe_sh}] give us realistic approximation for low $k$'s for linear emissivity. In the case of quadratic one self-convolution of spectra of such forms should be used instead:

\begin{equation} \label{eq:Fe_sq}
F_\eps(\vc{k}) 
  = \int F_\rho(\vc{k}') F_\rho(\vc{k}'-\vc{k}) \D{\vc{k}'}
\end{equation}

It is quite clear that for steep density the emissivity spectrum has asymptotically the same slope, than the underlying density one. However, for shallow density we have a different picture:

\begin{equation} \label{eq:Fe_sq_sh}
\begin{split} 
F_\eps(k) 
  & \sim \int
      \left( k_x'^2 + k_y'^2 + \left( k_z'- \frac{1}{2} k \right)^2 \right)^{-\frac{\alpha_\rho}{2}}
      \left( k_x'^2 + k_y'^2 + \left( k_z'+ \frac{1}{2} k \right)^2 \right)^{-\frac{\alpha_\rho}{2}}
    \D{\vc{k}'} \\
  & = 2\pi \int_0^\infty k'^2 \D{k'} \int_{-1}^{1}\left(
      \left( k'^2 + \frac{1}{4} k^2 \right)^2 - k'^2 k^2 t^2
    \right)^{-\frac{\alpha_\rho}{2}} \D{t} \\
  & \approx 4\pi \int_0^\infty \left( k'^2 + \frac{1}{4} k^2 \right)^{-\alpha_\rho}  k'^2 \D{k'} \\
  & \sim \dfrac{1}{k^{2\alpha_\rho-3}} \\
\end{split}
\end{equation}

Therefore, if emissivity is proportional to squared density, emissivity spectral index is as follows:

\begin{equation} \label{eq:aeps_sq}
\alpha_\eps
  = \left\{ 
    \begin{array}{ll} 
      \alpha_\rho, & \alpha_\rho \ge 3\\ 
      2\alpha_\rho-3, & \alpha_\rho < 3\\ 
    \end{array} 
  \right. 
\end{equation}

\section{Accounting for the linear velocity shear} \label{sect:reg_vel}  

Let us consider the case of ideal resolution. With account for Eq. [\ref{eq:wa_irp}] and keeping the regular velocity term, we will have Eq. [\ref{eq:P1_fin}] in the following form: 
\begin{equation} \label{eq:P1_ir_b} 
P_1(k_v)
  = \ft{f}^2(k_v) \int_{-\infty}^\infty w_{\eps,a}(z) \D{z} 
    C_\eps(z) \exp \left(-\frac{k_v^2}{2} D_z(z)-ik_v bz \right) 
\end{equation}  

Or, for high $k_v$'s, 
\begin{equation} \label{eq:P1_ir_b_asy} 
P_1(k_v) 
  \approx  
    \frac{ 
      2 w_{\eps,a} C_\eps(0) J_{z,1} 
    }{ 
      (2\pi V_0^2 I_c)^{\frac{1}{\alpha_v-3}}  
    } \cdot 
    \frac{\ft{f}^2(k_v)}{|k_v|^{\frac{2}{\alpha_v-3}}} \cdot G(k_v) 
\end{equation} 
\noindent where $G(k_v)$ is a factor, responsible for the linear velocity shear: 
\begin{equation} \label{eq:G_approx} 
G(k_v) 
  \approx \frac{1}{J_{z,1}} 
    \int_0^\infty \exp \left( -q^{\alpha_v-3} \right)  
    \cos \frac{q}{(H k_v)^{\frac{5-\alpha_v}{\alpha_v-3}}} \D{q} 
\end{equation} 
\noindent where 
\begin{equation} \label{eq:H} 
H 
  = (2\pi V_0^2 I_c)^{\frac{1}{5-\alpha_v}} 
    b^{-\frac{\alpha_v-3}{5-\alpha_v}} 
\end{equation} 
It is obvious, that 
\begin{equation*}  
\lim_{k_v \to \infty} G(k_v) = 1\,, \quad 3 < \alpha <5 
\end{equation*} 
\noindent so the high-$k_v$ asymptote stays valid in this case, too.

The same can be easily shown for the case of limited resolution.

\section{Constants, involved in $P_1$ asymptotics} \label{sect:Jxx}

\begin{equation} \label{eq:Jz_1} 
J_{z,1} 
  = \int_0^\infty \exp \left( -q^{\alpha_v-3} \right) \D{q}
\end{equation} 
\begin{equation} \label{eq:Jz_2} 
J_{z,2} 
  = \int_0^\infty \frac{\D{q}}{q^{3-\alpha_\eps}} \exp \left( -q^{\alpha_v-3} \right)
\end{equation} 
\begin{equation} \label{eq:Jr_1} 
J_{r,1} 
  = \int_0^\infty \exp \left( -q^{\alpha_v-3} \right) q^2 \D{q}
\end{equation} 
\begin{equation} \label{eq:Jth_1} 
J_{\theta,1} 
  = \int_0^\pi 
    \frac{ 
      \sin \theta \D{\theta} 
    }{ 
      (I_c \cos^2 \theta + I_s sin^2 \theta )^{\frac{3}{\alpha_v-3}}
    } 
\end{equation} 
\begin{equation} \label{eq:Jr_2} 
J_{r,2} 
  = \int_0^\infty \exp \left( -q^{\alpha_v-3} \right) q^{\alpha_\eps-1} \D{q}
\end{equation} 
\begin{equation} \label{eq:Jth_2} 
J_{\theta,2} 
  = \int_0^\pi 
    \frac{ 
      \sin \theta \D{\theta} 
    }{ 
      (I_c \cos^2 \theta + I_s sin^2 \theta )^{\frac{\alpha_\eps}{\alpha_v-3}}
    } 
\end{equation} 
\begin{equation} \label{eq:Jr_3} 
J_{r,3} 
  = \int_0^\infty \exp \left( -q^{\alpha_v-3} \right) q \D{q}
\end{equation} 
\begin{equation} \label{eq:Jth_3} 
J_{\theta,3} 
  = \int_0^\pi 
    \frac{ 
      \sin \theta \D{\theta} 
    }{
      \sqrt{2\pi \sin^2 \theta + 4\theta_0^2 \cos^2 \theta}
      (I_s sin^2 \theta + I_c \cos^2 \theta)^{\frac{2}{\alpha_v-3}}
    } 
\end{equation}
\begin{equation} \label{eq:Jr_4} 
J_{r,4} 
  = \int_0^\infty \exp \left( -q^{\alpha_v-3} \right) q^{\alpha_\eps-2} \D{q}
\end{equation} 
\begin{equation} \label{eq:Jth_4} 
J_{\theta,4} 
  = \int_0^\pi 
    \frac{ 
      \sin \theta \D{\theta} 
    }{
      \sqrt{2\pi \sin^2 \theta + 4\theta_0^2 \cos^2 \theta} 
      (I_s sin^2 \theta + I_c \cos^2 \theta)^{\frac{\alpha_\eps-1}{\alpha_v-3}}
    } 
\end{equation} 
\begin{equation} \label{eq:Jr_5} 
J_{r,5} 
  = \int_0^\infty \exp \left( -q^{\alpha_v-3} \right) q \D{q}
\end{equation} 
\begin{equation} \label{eq:Jth_5} 
J_{\theta,5} 
  = \int_0^\pi 
    \frac{ 
      \D{\theta} 
    }{
      (I_s sin^2 \theta + I_c \cos^2 \theta)^{\frac{2}{\alpha_v-3}}
    } 
\end{equation}
\begin{equation} \label{eq:Jr_6} 
J_{r,6} 
  = \int_0^\infty \exp \left( -q^{\alpha_v-3} \right) q^{\alpha_\eps-2} \D{q}
\end{equation} 
\begin{equation} \label{eq:Jth_6} 
J_{\theta,6} 
  = \int_0^\pi 
    \frac{ 
      \D{\theta} 
    }{
      (I_s sin^2 \theta + I_c \cos^2 \theta)^{\frac{\alpha_\eps-1}{\alpha_v-3}}
    } 
\end{equation} 
\begin{equation} \label{eq:Jr_7} 
J_{r,7} 
  = \int_0^\infty \exp \left( -q^{\alpha_v-3} \right) q \D{q} 
\end{equation} 
\begin{equation} \label{eq:Jth_7} 
J_{\theta,7} 
  = \int_0^\pi 
    \frac{ 
      \D{\theta} 
    }{
      (I_s sin^2 \theta + I_c \cos^2 \theta)^{\frac{2}{\alpha_v-3}}
    } 
    \log \left(
      \frac{
        \sqrt{\frac{1}{\theta_0} \sin^2 \theta + \cos^2 \theta }
      }{
        z_{edge} (2\pi V_0^2)^{\frac{1}{\alpha_v-3}} 
        (I_c \cos^2 \theta + I_s \sin^2 \theta)^{\frac{1}{\alpha_v-3}} 
        |k_v|^{\frac{2}{\alpha_v-3}}
      }
    \right)
\end{equation}
\begin{equation} \label{eq:Jr_8} 
J_{r,8}
  = \int_0^\infty \exp \left( -q^{\alpha_v-3} \right) q^{\alpha_\eps-2} \D{q}
\end{equation} 
\begin{equation} \label{eq:Jth_8} 
J_{\theta,8} 
  = \int_0^\pi 
    \frac{ 
      \D{\theta} 
    }{
      (I_s sin^2 \theta + I_c \cos^2 \theta)^{\frac{\alpha_\eps-1}{\alpha_v-3}}
    } 
    \log \left(
      \frac{
        \sqrt{\frac{1}{\theta_0} \sin^2 \theta + \cos^2 \theta }
      }{
        z_{edge} (2\pi V_0^2)^{\frac{1}{\alpha_v-3}} 
        (I_c \cos^2 \theta + I_s \sin^2 \theta)^{\frac{1}{\alpha_v-3}} 
        |k_v|^{\frac{2}{\alpha_v-3}}
      }
    \right)
\end{equation} 

Here the first index designates the integration domain.


\newpage

\begin{deluxetable}{llll}
\tablecaption{Asymptotics of $P_1(k_v)$ and $P_{1,v}(k_v)$ for $k_v \to \infty$ \label{tab:P1v_asy}}
\tablehead{
  \colhead{l.o.s. geometry} & \colhead{pencil beam} & \colhead{flat beam} & \colhead{low resolution}
}
\startdata
parallel
&
\begin{math} 
  \frac{A_v \ft{f}^2(k_v)}{|k_v|^{\frac{2}{\alpha_v-3}}}
\end{math}
&
\begin{math} 
  \frac{A_v \ft{f}^2(k_v)}{|k_v|^{\frac{4}{\alpha_v-3}}} 
\end{math}
&
\begin{math}
  \frac{A_v \ft{f}^2(k_v)}{|k_v|^{\frac{6}{\alpha_v-3}}} 
\end{math}
\\
crossing 
&
\begin{math} 
  \frac{A_v \ft{f}^2(k_v)}{|k_v|^{\frac{2}{\alpha_v-3}}}
\end{math}
&
\begin{math} 
  \frac{
    A_v \ft{f}^2(k_v) \log \left( \left|\frac{k_v}{k_{v,f}}\right|^{\frac{2}{\alpha_v-3}} \right) 
  }{
    |k_v|^{\frac{4}{\alpha_v-3}
  }
} 
\end{math}
&
\begin{math} 
  \frac{A_v \ft{f}^2(k_v)}{|k_v|^{\frac{4}{\alpha_v-3}}} 
\end{math}
\\
\enddata
\tablecomments{
  Amplitudes $A_v$ see in Tab. \ref{tab:Au}. 
}
\end{deluxetable}

\begin{deluxetable}{llll}
\tablecaption{Normalization coefficients $A_v$ for $P_1(k_v)$ and $P_{1,v}(k_v)$ \label{tab:Au}}
\tablehead{
  \colhead{l.o.s. geometry} & \colhead{pencil beam} & \colhead{flat beam} & \colhead{low resolution}
}
\startdata
parallel
&
\begin{math} 
  \frac{ 
   2 w_{\eps,a}(0) \sigma_\eps J_{z,1} 
  }{ 
    \left( 2\pi V_0^2 I_c \right)^{\frac{1}{\alpha_v-3}}  
  } 
\end{math}
&
\begin{math} 
  \frac{ 
      W_{b,\theta,a}(0) w_{\eps,a}(0) \sigma_\eps J_{r,5} J_{\theta,5} 
    }{ 
      (2\pi V_0^2)^{\frac{2}{\alpha_v-3}}  
  }
\end{math}
&
\begin{math}
  \frac{ 
      2 \pi w_a(\vc{0}) \sigma_\eps J_{r,1} J_{\theta,1} 
  }{ 
    (2\pi V_0^2)^{\frac{3}{\alpha_v-3}}  
  }
\end{math}
\\
crossing 
&
\begin{math} 
  \frac{ 
   2 w_{\eps,a}(0) \sigma_\eps J_{z,1} 
  }{ 
    \left( 2\pi V_0^2 I_c \right)^{\frac{1}{\alpha_v-3}}  
  } 
\end{math}
&
\begin{math} 
  \frac{
    \sigma_\eps J_{r,7} J_{\theta,7}
  }{
    \sqrt{2\pi} \theta_0 (2\pi V_0^2)^{\frac{2}{\alpha_v-3}}
  }
\end{math}
&
\begin{math} 
  \frac{ 
    \pi \sigma_\eps J_{r,3} J_{\theta,3} 
  }{ 
    \theta_0 (2\pi V_0^2)^{\frac{2}{\alpha_v-3}}  
  }
\end{math}
\\
\enddata
\tablecomments{
  For $P_1$ $\sigma_\eps = C_\eps(\vc{0})$, for $P_{1,v}$ $\sigma_\eps = \avg{\eps}^2$.
}
\end{deluxetable}

\begin{deluxetable}{llll}
\tablecaption{Asymptotics of $P_{1,\eps}(k_v)$ for $k_v \to \infty$ \label{tab:P1eps_asy}}
\tablehead{
  \colhead{l.o.s. geometry} & \colhead{pencil beam} & \colhead{flat beam} & \colhead{low resolution}
}
\startdata
parallel
&
\begin{math} 
  \frac{A_\eps \ft{f}^2(k_v)}{|k_v|^{2 \frac{\alpha_\eps-2}{\alpha_v-3}}} 
\end{math}
&
\begin{math} 
  \frac{A_\eps \ft{f}^2(k_v)}{|k_v|^{2 \frac{\alpha_\eps-1}{\alpha_v-3}}} 
\end{math}
&
\begin{math}
  \frac{A_\eps \ft{f}^2(k_v)}{|k_v|^{\frac{2 \alpha_\eps}{\alpha_v-3}}} 
\end{math}
\\
crossing 
&
\begin{math} 
  \frac{A_\eps \ft{f}^2(k_v)}{|k_v|^{2 \frac{\alpha_\eps-2}{\alpha_v-3}}} 
\end{math}
&
\begin{math} 
  \frac{
    A_\eps \ft{f}^2(k_v) \log \left( \left|\frac{k_v}{k_{v,f}}\right|^{\frac{2}{\alpha_v-3}} \right) 
  }{
    |k_v|^{2 \frac{\alpha_\eps-1}{\alpha_v-3}}
  } 
\end{math}
&
\begin{math} 
  \frac{A_\eps \ft{f}^2(k_v)}{|k_v|^{2 \frac{\alpha_\eps-1}{\alpha_v-3}}} 
\end{math}
\\
\enddata
\tablecomments{
  Amplitudes $A_\eps$ see in Tab. \ref{tab:Aeps}.
}
\end{deluxetable}

\begin{deluxetable}{llll}
\tablecaption{Normalization coefficients $A_\eps$ for $P_{1,\eps}(k_v)$ \label{tab:Aeps}}
\tablehead{
  \colhead{l.o.s. geometry} & \colhead{pencil beam} & \colhead{flat beam} & \colhead{low resolution}
}
\startdata
parallel
&
\begin{math}
  \frac{
    8\pi w_{\eps,a}(0) \mcl{E}_0^2 I_{sh}^\infty J_{z,2} 
  }{
    \left( 2\pi V_0^2 I_c \right)^{\frac{\alpha_\eps-2}{\alpha_v-3}}  
  }
\end{math}
&
\begin{math} 
  \frac{ 
    2 \pi W_{b,\theta,a}(0) w_{\eps,a}(0) \mcl{E}_0^2 I_{sh}^\infty J_{r,6} J_{\theta,6} 
  }{ 
    (2\pi V_0^2)^{\frac{\alpha_\eps-1}{\alpha_v-3}}  
  }
\end{math}
&
\begin{math}
  \frac{ 
    8\pi^2 w_a(\vc{0}) \mcl{E}_0^2 I_{sh}^\infty J_{r,2} J_{\theta,2} 
  }{ 
    (2\pi V_0^2)^{\frac{\alpha_\eps}{\alpha_v-3}}  
  }
\end{math}
\\
crossing 
&
\begin{math} 
  \frac{ 
    8\pi w_{\eps,a}(0) \mcl{E}_0^2 I_{sh}^\infty J_{z,2} 
  }{ 
    \left( 2\pi V_0^2 I_c \right)^{\frac{\alpha_\eps-2}{\alpha_v-3}}  
  }
\end{math}
&
\begin{math} 
  \frac{ 
    2 \pi \mcl{E}_0^2 I_{sh}^\infty J_{r,8} J_{\theta,8} 
  }{ 
    (2\pi V_0^2)^{\frac{\alpha_\eps-1}{\alpha_v-3}}  
  }
\end{math}
&
\begin{math} 
  \frac{ 
    4\pi^2 \mcl{E}_0^2 I_{sh}^\infty J_{r,4} J_{\theta,4} 
  }{ 
    (2\pi V_0^2)^{\frac{\alpha_\eps-1}{\alpha_v-3}}  
  }
\end{math}
\\
\enddata
\end{deluxetable}

\begin{deluxetable}{llll}
\tablecaption{Asymptotics of geometric term $w_a(\vc{r})$ for $r \to 0$ \label{tab:gc}}
\tablehead{
  \colhead{l.o.s. geometry} & \colhead{pencil beam} & \colhead{flat beam} & \colhead{low resolution}
}
\startdata
parallel
&
\begin{math} 
w_{\eps,a}(0) \delta(\vc{R})
\end{math}
&
\begin{math} 
w_{\eps,a}(0) W_{b,\theta,a}(0) \delta(x)
\end{math}
&
\begin{math} 
w_{\eps,a}(0) W_{b,a}(0) 
\end{math}
\\
crossing 
&
\begin{math} 
w_{\eps,a}(0) \delta(\vc{R})
\end{math}
&
\begin{math} 
  - \frac{\delta(x)}{\sqrt{2\pi} \theta_0} 
  \log \left( \frac{r}{z_{edge}} \sqrt{\frac{1}{\theta_0} \sin^2 \theta + \cos^2 \theta }\right)
\end{math}
&
\begin{math} 
  \frac{1}{2 \theta_0 r \sqrt{2 \pi \sin^2 \theta + 4 \theta_0^2 \cos^2 \theta}}
\end{math}
\\
\enddata
\tablecomments{
  $\vc{r} \equiv (x,y,z)$, $\vc{R} \equiv (x,y)$, $\theta$ is an angle between $\vc{r}$ and $\hat{z}$.
}
\end{deluxetable}

\begin{deluxetable}{lll}
\tablecaption{List of symbols\label{tab:Symb}}
\tablehead{
  \colhead{symbol} & \colhead{description} & \colhead{defined at}
}
\startdata
$\vc{r}$ & radius-vector & Sect. \ref{sect:SLine}  \\
$\vc{k}$ & Fourier counterpart of $\vc{r}$ & Appx. \ref{sect:vst} \\
$\vc{R}$ & picture-plain projection of $\vc{r}$ & Sect. \ref{sect:CFS} \\
$v$ & velocity &  Sect. \ref{sect:SLine} \\
$k_v$ & Fourier counterpart of $v$ &  Sect. \ref{sect:SLine}  \\
$\alpha_v$, $\alpha_\eps$, $\alpha_\rho$ & spectral indices for velocity, emissivity and density & [\ref{eq:Fu}], [\ref{eq:Fe_st}], Appx. \ref{sect:ecf_sq}\\
$n_0$ & number density & [\ref{eq:dens}] \\
$\varphi(v)$ & Maxwellian distribution & [\ref{eq:maxw}] \\
$\eps_0$, $\eps$ & emissivity, real and idealized (homogeneous) & [\ref{eq:emiss}], [\ref{eq:eps0}] \\
$S$ & spectral line &  [\ref{eq:sig}] \\
$f_s$, $f$ & channel sensitivity function, original and effective &  [\ref{eq:sig}], [\ref{eq:effsens}] \\
$\lambda$ & wavelength of a spectral line &  [\ref{eq:sig}] \\
$w_b$ & instrument beam & [\ref{eq:sig}] \\
$w_\eps$ & factor, defining the observed object boundaries  & [\ref{eq:eps0}] \\
$w$ & window function &  [\ref{eq:w}] \\
$K$ & statistics, common for for VCA and VCS &  [\ref{eq:K12_def}] \\
$D_z$ & velocity structure tensor projection &  [\ref{eq:Dz_def}] \\
$C_\eps$ & emissivity correlation function &  [\ref{eq:K12_inter}] \\
$w_{12}$, $w_a$, $w_{\eps,a}$ & geometric term: general, for coinciding beams, for $w_\eps$ &  [\ref{eq:w12_def}], [\ref{eq:wa_def}], [\ref{eq:w_eps_a_z}] \\
$P_1$ & power spectrum along velocity coordinate & [\ref{eq:P1_fin}] \\
$W_b$ & auxiliary function, derived from $w_b$ & [\ref{eq:Wb_plos_def}], [\ref{eq:Wb_clos_def}] \\
$W_{b,a}$ & auto-convolution of $W_b$ &  [\ref{eq:Wba_plos}] \\
$I_c$, $I_s$ & auxiliary functions for $D_z$ &  [\ref{eq:Ic}], [\ref{eq:Is}] \\
$J_{*,*}$ & auxiliary constants & Appx. \ref{sect:Jxx} \\
\enddata
\tablecomments{
Tilde means Fourier transform, $v$ and $\eps$ subscripts in $P_1$ and $K$ mean velocity and emissivity terms according to Eq. [\ref{eq:P1_split}].
}
\end{deluxetable}


\end{document}